\documentclass[fleqn,usenatbib]{mnras}

\usepackage{newtxtext,newtxmath}

\usepackage[T1]{fontenc}
\usepackage{ae,aecompl}

\usepackage{graphicx}	
\usepackage{amsmath}	
\usepackage{amssymb}	

\title[Overcoming the structural surface effect]{Overcoming the structural surface effect with a realistic treatment of turbulent convection in 1D stellar models}

\author[A. C. S. J{\o}rgensen \& A. Weiss]{
Andreas Christ S{\o}lvsten J{\o}rgensen$^{1}$\thanks{E-mail: acsj@mpa-garching.mpg.de},
Achim Weiss$^{1}$
\\
$^{1}$Max-Planck-Institut f\"ur Astrophysik, Karl-Schwarzschild-Str. 1, D-85748 Garching, Germany
}

\date{Accepted 12.07.2019. Received 09.07.2019; in original form 08.03.2019}

\pubyear{2019}

\begin{document}
\label{firstpage}
\pagerange{\pageref{firstpage}--\pageref{lastpage}}
\maketitle

\begin{abstract}
State-of-the-art one-dimensional (1D) stellar evolution codes rely on simplifying assumptions, such as mixing length theory, in order to describe superadiabatic convection. As a result, 1D stellar structure models do not correctly recover the surface layers of the Sun and other stars with convective envelopes. We present a method that overcomes this structural drawback by employing three-dimensional (3D) hydrodynamic simulations of stellar envelopes: at every time-step of the evolution interpolated 3D envelopes are appended to the 1D structure and are used to supply realistic boundary conditions for the stellar interior. In contrast to previous attempts, our method includes mean 3D turbulent pressure. We apply our method to model the present Sun. The structural shortcomings of standard stellar models lead to systematic errors in the stellar oscillation frequencies inferred from the model. We show that our method fully corrects for this error. Furthermore, we show that our realistic treatment of superadiabatic convection alters the predicted evolution of the Sun. Our results hence have important implications for the characterization of stars. This has ramifications for neighbouring fields, such as exoplanet research and galactic archaeology, for which accurate stellar models play a key role.
\end{abstract}

\begin{keywords}
Asteroseismology -- stars: interiors -- stars: atmospheres
\end{keywords}



\section{Introduction} \label{sec:intro}

Oscillation frequencies computed from standard one-dimensional (1D) stellar structure models show, among other deficits, a systematic deviation from observations that increases with increasing frequency. This discrepancy is known as the surface effect \citep{Brown1984,Christensen-Dalsgaard1988} and results from an inadequate modeling of the stellar atmosphere and of the superadiabatic convective surface layers: usually, superadiabatic convection is modelled using simplifying approximations, such as mixing length theory \citep{Bohm-Vitense1958} or full spectrum turbulence theories \citep{Canuto1991,Canuto1992}. Furthermore, artificial atmospheres, such as a plane-parallel Eddington-grey one, are used to set the outer boundary conditions. With the high-quality data from the CoRoT \citep{Baglin2009} and \textit{Kepler} \citep{Borucki2010x} space missions, the surface effect has become a crucial limitation, since seismic measurements have become a common tool to constrain stellar parameters \citep[e.g.][]{jcd2010,Chaplin2013}. 

One way to mitigate the surface effect is to correct the predicted frequencies by empirical or semi-empirical relations \citep[see, for example,][]{Kjeldsen2008,Ball2014,Sonoi2015}. An alternative approach is to improve the stellar models themselves by replacing the incorrect superadiabatic layers by a structure obtained from multi-dimensional radiative hydrodynamic (RHD) simulations. In this connection, many authors have followed the so-called patching approach, where the substitution is performed in the post-processing: the stellar model is evolved in the usual way, but the outermost layers of the final model are replaced by the temporally and spatially averaged structure of a multi-dimensional RHD simulation. In the following, we will refer to these mean 3D structures as $\langle \mathrm{3D} \rangle$-envelopes.
The oscillation frequencies of the resulting patched models show an improved agreement with observations. Examples for such investigations are \citet{Piau2014,Sonoi2015,Ball2016,Magic2016,Joergensen2017,Trampedach2017,Joergensen2019}.

Patching was initially restricted to stellar models with global parameters that were identical to those of existing 3D simulations. Until recently, most studies were hence limited to the present Sun. This changed with the publication of an interpolation scheme by \cite{Joergensen2017}: the interpolation scheme is capable of accurately predicting the stratification of $\langle \mathrm{3D} \rangle$-envelopes for any combination of effective temperature ($T_\mathrm{eff}$) and gravitational acceleration ($g$). The accuracy of the scheme was comprehensively tested using cross validation, i.e. by withholding individual envelopes and showing that the scheme reliably recovered these based on the remaining 3D simulations. For further details, we refer to \cite{Joergensen2017,Joergensen2019}. 
{\color{black}In \cite{Joergensen2017}, this interpolation scheme was limited to one composition.}
The scheme was recently extended upon by \cite{Joergensen2019}, allowing for an interpolation in metallicity ($\mathrm{[Fe/H]}$).

Since the atmosphere and the superadiabatic outer layers provide the boundary conditions for the stellar interior, the use of MLT and grey atmospheres leads to inaccurate radii and effective temperatures of stellar models. Inadequate boundary layers hence affect the predicted stellar evolution tracks, which patching does not correct for: patching only corrects for the structural shortcomings and the associated surface effect of the final model.
In order to address this issue, \cite{Joergensen2018} suggested an alternative approach, building upon the interpolation scheme by \cite{Joergensen2017}: at every time-step of the evolution, interpolated $\langle \mathrm{3D} \rangle$-envelopes are used as the outer layers and to set the boundary conditions for the interior structure. \cite{Joergensen2018} applied the method to model the present Sun. The resulting 1D standard solar model (SSM) was shown to correctly reproduce the mean stratification of 3D simulations, which strongly reduced the structural shortcomings that contribute to the surface effect. The remaining structural inadequacies of the surface layers of these models can mostly be attributed to the neglect of turbulent pressure in the hydrostatic structure. In the following we will refer to this approach as the \textit{coupling of 1D and 3D simulations}.

In this paper, we extend upon the method presented by \cite{Joergensen2018} by including turbulent pressure. Thus, at each time-step, we include turbulent pressure in the appended $\langle \mathrm{3D} \rangle$-envelopes. Below the outer boundary of the interior model, we furthermore scale the turbulent pressure as estimated from MLT, in order to achieve a {\color{black}continuous} transition between the appended $\langle \mathrm{3D} \rangle$-envelope and the stellar interior. As shown in Section~\ref{sec:completemodel}, we are hereby able to fully eliminate the structural contribution to the surface effect. The predicted model frequencies hence only suffer from modal effects, such as non-adiabatic energetics. In other words, the remaining discrepancy between the inferred oscillations and the observed p-mode frequencies solely reflect simplifying assumptions that enter the frequency computation, while the underlying stellar structure is correct \citep[e.g.][]{Houdek2017}.


\section{Standard 1D stellar models} \label{sec:standard}

In this section, we give a short introduction to state-of-the art standard stellar models, which include the MLT-theory for the outermost superadiabatic convective layers and simplified atmospheres. 

The use of MLT and simplistic model atmospheres has been shown to be incapable of recovering the stratification of more realistic simulations \citep{Schlattl1997,Trampedach2011, Joergensen2017,Mosumgaard2018}.
In Section~\ref{sec:Coupling1D3D}, we will discuss how to remedy this shortcoming by including information from 3D simulations. By introducing standard stellar structure models below, we elaborate upon concepts that we will draw upon in the following sections. Also, we will compare the results obtained from coupling 1D and 3D simulations with those obtained from standard 1D structure models.

Throughout this paper we employ \textsc{garstec} \citep[Garching Stellar Evolution Code;][]{Weiss2008}, a standard one-dimensional (1D) stellar evolution code assuming spherical symmetry.
Computing stellar structure and evolution models amounts to solving the stellar structure equations, a set of coupled differential equations, for given mass $M$ and composition. In \textsc{garstec}, these equations are solved employing the Henyey-scheme \citep{Henyey1964}: at each time-step the structure is iteratively corrected, until the stellar structure equations are fulfilled with the required degree of accuracy at every mesh point of the 1D model.

Having established the equilibrium structure at a given time-step, the chemical profile is evolved in time based on the relevant nuclear reactions and any particle transport effect such as convective mixing or atomic diffusion. During the computation of the chemical evolution, the temperature and pressure stratifications are kept fixed. Subsequently, a new equilibrium structure is determined, using the Henyey scheme. This is repeated for every time-step, alternating between the computation of an equilibrium structure and the change in the chemical profile. In this manner, the evolution of the structure along the evolutionary track is determined. For a comprehensive overview of the Henyey scheme and the details of the explicit time integration in \textsc{garstec} we refer to \cite{Schlattl1997,Kippenhahn}.

In order to solve the stellar structure equations to obtain a new equilibrium structure, suitable boundary conditions must be supplied {\color{black}both at the centre and the surface --- in this paper, we are not concerned with the inner boundary conditions and will not address these further}. In the case of standard stellar structure models, \textsc{garstec} supplies the {\color{black}outer} boundary conditions at the stellar surface, i.e. the photosphere: first, Stefan-Boltzmann's law is required to be fulfilled:
\begin{equation}
L = 4 \pi \sigma R^2 T_\mathrm{eff}^4. \label{eq:SBlaw}   
\end{equation}
Here, $\sigma$ denotes the Stefan-Boltzmann constant, $R$ is the surface radius, $L$ is the total luminosity, and $T_\mathrm{eff}$ is the effective temperature. Eq.~(\ref{eq:SBlaw}) follows from the definition of the effective temperature based on the surface energy flux ($F_\mathrm{tot}$) of the star
\begin{equation}
F_\mathrm{tot} = \sigma T_\mathrm{eff}^4 \label{eq:flux},
\end{equation}
and from the connection between luminosity and flux. The definition of the surface radius depends on the employed model atmosphere: for the Eddington grey atmosphere discussed below, the photosphere is located at an optical depth of $2/3$ \citep{Kippenhahn}. When constructing coupled models, we define the surface of the model to be located at the radius, at which the temperature equals the attributed $T_\mathrm{eff}$ (cf.~Section~\ref{sec:Coupling1D3D}).

Second, the temperature and pressure at the photosphere is required to match the results obtained by integration of an model atmosphere that sets the outer boundary conditions. In this paper, we use an Eddington grey atmosphere, when computing standard stellar structure models. For such atmospheres, the temperature stratification is assumed to be 
\begin{equation}
T^4(\tau) = \frac{3}{4}\frac{L}{4\pi R^2\sigma}\left( \tau + \frac{2}{3} \right), \label{eq:Eddington}
\end{equation}
where $T$ denotes the temperature, $P$ denotes the pressure, and $\tau$ denotes the optical depth. By using Eq.~(\ref{eq:Eddington}), we assume the atmosphere to be fully radiative{\color{black}. We note that alternative model atmospheres, including {\color{black}semi-empirical approximations \citep[e.g.][]{KrishnaSwamy1966} as well as more realistic theoretical model atmospheres \citep[e.g.][]{Schlattl1997,2008A&A...486..951G,Trampedach2014a}}, are available. For these alternative atmosphere models, the optical depth at the photosphere differs from 2/3.

The interior model is computed, assuming the diffusion approximation for radiative transport to be valid. This approximation holds true, if the gas is optically thick. Meanwhile, the optically thin atmosphere must solve the radiative transfer equation. The Eddington grey atmosphere is an approximation that implies that the transition between the optically thin and thick regions takes place at $\tau=2/3$, whilst this transition, realisitically, {\color{black}is not completed until} $\tau \gtrapprox 10$. This deficit is usually ignored. In other words, standard stellar models give an incomplete description of the photospheric transition. When coupling of 1D and 3D models, on the other hand, we overcome this shortcoming, by applying the boundary conditions much deeper within the stellar structure (cf.~Section~\ref{sec:Coupling1D3D}). 
}

The differential equations that provide the pressure and the temperature depend on other quantities, such as the density ($\rho$) or the opacity. We use the OPAL opacities \citep{Iglesias1996} in combination with the opacities by \cite{Ferguson2005} for low temperatures. 
Like several other quantities, including the first adiabatic index ($\Gamma_1$) and the entropy, the density is derived from the equation of state (EOS). Here we employ the FreeEOS by A.~W. Irwin \citep{Cassisi2003}. The values derived from the EOS also depend on the chemical composition, here being the solar composition by \cite{Asplund2009}, to which we will refer as AGSS09. This chemical composition is chosen in order to be consistent with the choices made for the 3D simulations of convective envelopes ({\color{black}cf.} Section~\ref{sec:stagger}). This enforced consistency facilitates a straight-forward comparison between the various solar calibration models.


\section{The Stagger grid} \label{sec:stagger}

For the 3D RHD models, we draw upon the so-called ``Stagger grid'' by \cite{Magic2013}. Each of these simulations covers a representative volume of a stellar envelope, stretching from the atmosphere above the photosphere to the nearly adiabatic interior. Thus, the depth of each envelope only corresponds $0.4-10\,\%$ of the stellar radius \citep[cf.][]{Magic2013}. Due to the small extent of the envelopes, the gravitational acceleration is set to be constant within each simulation.

The Stagger grid contains 206 models, spanning effective temperatures between $4000\,$K and $7000\,K$ in steps of $500\,$K. The logarithm of the gravitational acceleration ($g$) varies between $5.0\,$dex and $1.5\,$dex in steps of $0.5\,$dex. Thus, the Stagger grid includes stellar envelopes of low-mass stars from the main-sequence (MS) to the red giant branch (RGB). The grid spans simulations with metallicities, i.e. $\mathrm{[Fe/H]}$, of $-4.0$, $-3.0$, $-2.0$, $-1.0$, $-0.5$, $0.0$ and $0.5$. The Stagger grid employs the EOS by \cite{Hummer1988}, and the composition is based on AGSS09.

{\color{black}The Stagger-grid simulations extend from very low optical depths to optically thick and convective regions below the photosphere. Whilst they are rather shallow compared to the entire convection zone, they nevertheless cover most of the superadiatic outermost layers {\color{black}--- thus, when we refer to the outermost layers, we refer to the region that is covered by the 3D simulations.}
In order emphasize this and to distinguish them from the purely radiative grey atmospheres that are commonly employed in standard 1D models, we will refer to them as $\langle \mathrm{3D} \rangle$-\textit{envelopes} rather than $\langle \mathrm{3D} \rangle$-atmospheres. In doing so, we adopt the nomenclature by \citet{Joergensen2018}.
}


\subsection{Interpolation} \label{sec:interp}

In this paper, we use the interpolation method by \cite{Joergensen2017}, which allows for an interpolation in the effective temperature ($T_\mathrm{eff}$) and gravitation acceleration ($g$). We do not perform any interpolation in the chemical composition and hence restrict ourselves to 28 simulations at solar metallicity. Since we present solar calibration models, this is a permissible simplification. 

The interpolation scheme by \cite{Joergensen2017} relies on the apparent homology of $\langle \mathrm{3D} \rangle$-envelopes: the stratifications of all relevant quantities are similar across {\color{black}the covered region of} the HR diagram, when scaling each quantity by the corresponding value at a fixed location defined by the run of density with pressure or radius. In red giants there is a density inversion due to the strong ionization of H and He \citep{Schwarzschild1975}, which in dwarfs still appears as a plateau. This is the reference point for the scaling, to which we will refer for simplicity as the ``density inversion''.
Before the interpolation, all 28 Stagger-grid simulations are hence scaled by the corresponding values at the minimum in $\partial \log \rho / \partial \log P_\mathrm{th}$. Here, $\rho$ denotes the density, while $P_\mathrm{th}$ denotes the thermal pressure, i.e. the sum of the gas pressure ($P_\mathrm{gas}$) and the radiation pressure ($P_\mathrm{rad}$):
\begin{equation}
P_\mathrm{th}=P_\mathrm{gas}+P_\mathrm{rad}.
\end{equation}
Having determined the scaled structures of the Stagger-grid simulations, we interpolate, in order to obtain the scaled structure and the associated scaling factors at the desired $T_\mathrm{eff}$ and $g$. The scaled thermal pressure is used as the coordinate. Having obtained the scaled structure, the scaling is inverted. For further details on the interpolation scheme, we refer to \cite{Joergensen2017}.


\section{Coupling 1D and 3D simulations} \label{sec:Coupling1D3D}

In this paper, we employ and extend upon the method developed by \cite{Joergensen2018}, in order to include interpolated $\langle \mathrm{3D} \rangle$-envelopes in our stellar evolution code. Thus, we append $\langle \mathrm{3D} \rangle$-envelopes at every time-step and use the base of the appended $\langle \mathrm{3D} \rangle$-envelope to set the boundary conditions of the interior model.

{\color{black}Note that the procedure described below shows many similarities with the standard procedure. We hence draw upon several concepts that were introduced in Section~\ref{sec:standard}.}

The coupled stellar structure models are determined, using the Henyey scheme (cf. Section~\ref{sec:standard}). In each iteration, the solution of the stellar structure equations {\color{black}returns} a given temperature and thermal pressure at the outermost point of the interior model. In the following, we will use the nomenclature introduced by \cite{Joergensen2018} and we will hence refer to these values of temperature and pressure as $T^{\mathrm{1D}}_\mathrm{m}$ and $P^{\mathrm{1D}}_\mathrm{th,m}$, respectively. The subscript 'm' indicates that the values are taken at the outer boundary of the interior model, which, contrary to the standard model, no longer needs to be at the photosphere, but will be located in the lower regions of the superadiabatic convective envelope. Like \cite{Joergensen2018}, we will occasionally refer to the base of the appended envelope as the matching point, which motivates the use of the letter 'm'.

In each iteration, we compute and append an interpolated $\langle \mathrm{3D} \rangle$-envelope that by construction has the same temperature ($T^{\mathrm{3D}}_\mathrm{m}$) at its base as the interior model requires --- that is, $T^{\mathrm{3D}}_\mathrm{m} = T^{\mathrm{1D}}_\mathrm{m}$. This requirement is merely imposed in order to ensure that the temperature stratification is continuous at the transition from the interior model to the appended envelope.

The scaled thermal pressure at the base of the appended $\langle \mathrm{3D} \rangle$-envelope is consistently 16 times higher than at the density inversion near the surface. {\color{black}We refer to Mosumgaard et al. (in preparation) for a detailed discussion on which scaled pressure to use}. For the present-day Sun, this means that we set the outer boundary conditions of the interior model roughly 1 Mm below the photosphere, where the entropy is approximately constant \citep[cf.][]{Joergensen2018}. Thus, the outer boundary of the interior model is located in layers deep enough such that the convective stratification is nearly adiabatic. {\color{black}Most of the} superadiabatic layers of coupled models, {\color{black}including the photosphere}, are enclosed in the appended $\langle \mathrm{3D} \rangle$-envelope.
In contrast, when computing stellar models using the standard procedure described in Section~\ref{sec:standard}, the outer boundary conditions are supplied at the photosphere. For standard stellar models, the outer superadiabatic layers are therefore a part of the interior model, whilst this is not the case for the coupled models.

{\color{black}As already described in Section~\ref{sec:interp}, we compute $\langle \mathrm{3D} \rangle$-envelopes by interpolation in two global parameters:  $\log g$ and $T_\mathrm{eff}$. To perform the interpolation, propositions for $\log g$ and $T_\mathrm{eff}$ are needed. In this connection, we require the surface gravity of the $\langle \mathrm{3D} \rangle$-envelope to match the surface gravity at the outer boundary of the interior model.} The surface gravity of the interpolated $\langle \mathrm{3D} \rangle$-envelope is hence dictated by the mass and radius of the interior model at the {\color{black}matching point. In other words, for any given iteration, the proposition for $\log g$ comes directly from the proposed interior structure.}

For any given gravitational acceleration, a unique $T_\mathrm{eff}$ fulfills the requirement that $T^{\mathrm{3D}}_\mathrm{m} = T^{\mathrm{1D}}_\mathrm{m}$ at the required scaled pressure. Note that the resulting $T_\mathrm{eff}$ is merely a proposition for the current iteration: as specified below other conditions must likewise be fulfilled. Notably, $T_\mathrm{eff}$ must be consistent with the total energy flux of the coupled model in order to comply to its definition.
If these conditions are not fulfilled, the Henyey scheme will alter the interior structure, leading to a new $T^{\mathrm{1D}}_\mathrm{m}$, which results in a new proposition for $T_\mathrm{eff}$ {\color{black}and $\log g$}.

Having proposed $T_\mathrm{eff}$ based on the required match in temperature, the scaling factor for the thermal pressure can be {\color{black}obtained} by interpolation, as shown by \citet{Joergensen2017}, where it was found that the scaling factor varies rather linearly with the global stellar parameters. The scaled thermal pressure can hence be converted to the absolute thermal pressure ($P^{\mathrm{3D}}_\mathrm{th,m}$).

From the appended $\langle \mathrm{3D} \rangle$-envelope, only the turbulent pressure (cf.~Section~\ref{sec:turb}) and the temperature stratification as a function of the thermal pressure is obtained from the Stagger grid by interpolation. Other quantities, such as the density ($\rho$) or the first adiabatic index ($\Gamma_1$), are evaluated from the EOS of the 1D stellar evolution code. The chemical composition of the appended, convective $\langle \mathrm{3D} \rangle$-envelope is set to be the same as at the outermost point of the interior model, since the convective envelope is fully mixed. 

{\color{black}The chemical composition of the appended $\langle \mathrm{3D} \rangle$-envelope is hence consistent with the composition of the interior model throughout the computed evolution. However, since we do not interpolate in metallicity, we note that this composition will only be fully consistent with the metallicity of the underlying Stagger-grid simulations for the present-day Sun: as mentioned above, we have restricted ourselves to interpolation in 3D simulations, for which $\mathrm{[Fe/H]}=0.0$. 
}

The stellar model is now subject to two outer boundary conditions. Firstly, the stellar structure equations must recover the interpolated pressure at the base of the appended $\langle \mathrm{3D} \rangle$-envelope, i.e. $P^{\mathrm{1D}}_\mathrm{th,m} = P^{\mathrm{3D}}_\mathrm{th,m}$. This ensures that the pressure stratification is continuous at the base of the appended envelope.
Secondly, the Stefan-Boltzmann law, i.e.\  Eq.~(\ref{eq:SBlaw}), must be fulfilled.
Assuming that any energy contribution from the appended $\langle \mathrm{3D} \rangle$-envelope is negligible, we use the luminosity ($L_\mathrm{m}^\mathrm{1D}$) at the outermost point of the interior model to evaluate this criterion.
We use the proposed $T_\mathrm{eff}$ from the appended envelope {\color{black}found by interpolation}.

To obtain the surface radius, we proceed as follows. The radius ($r$) of each grid point within the appended $\langle \mathrm{3D} \rangle$-envelope is calculated from the requirement of hydrostatic equilibrium:
\begin{equation}
\frac{\mathrm{d} P}{\mathrm{d} r} = - \frac{Gm}{r^2}\rho = - \rho g.  \label{eq:hydrostat}
\end{equation}
Here, $G$ denotes the gravitational constant, $m=M(r)$ is the mass interior to $r$, and $P$ is the total pressure:
\begin{equation}
P=P_\mathrm{th}+P_\mathrm{turb} \label{eq:ptot},
\end{equation}
where $P_\mathrm{turb}$ denotes the turbulent pressure. Having obtained the radius of each mesh point in the model, the surface radius $R$ that enters the Stefan-Boltzmann law can be computed and the second boundary condition can be checked: we define the surface radius as the distance from the centre, at which the temperature equals $T_\mathrm{eff}$ --- that is, the determination of the surface radius involves interpolation in the appended envelope.

We note that \cite{Joergensen2018} solely employ the gas pressure, when computing interpolated $\langle \mathrm{3D} \rangle$-envelopes. By using the thermal pressure as the coordinate for the performed interpolation, we have hence added the radiation pressure to the implementation of $\langle \mathrm{3D} \rangle$-envelopes in \textsc{garstec}. 

Note that the described procedure for constructing coupled models shows many similarities to the standard procedure discussed in Section~\ref{sec:standard}: {\color{black}in both cases, the interior structure, including the stratification of all thermodynamic quantities, is iteratively adjusted, using the Henyey scheme. The stratification beyond the outer boundary of the interior model, i.e. for the Eddington atmosphere or the $\langle \mathrm{3D} \rangle$-envelope, is likewise iteratively altered. Furthermore, in both cases, the Stefan-Boltzmann law and the requirement to obtain continuous transitions in all parameters at the outer boundary are used to set the outer boundary conditions. The only notable difference}
is that the photosphere, at which the Stefan-Boltzmann law must be fulfilled, no longer corresponds to the outer boundary of the interior model, where a continuous transition in $T$ and $P$ is imposed.

In summary, in the case of coupled models, a consistent equilibrium structure must fulfill four conditions: the surface gravity of the appended envelope must be consistent with the interior model, $T^{\mathrm{3D}}_\mathrm{m} = T^{\mathrm{1D}}_\mathrm{m}$, $P^{\mathrm{3D}}_\mathrm{m} = P^{\mathrm{1D}}_\mathrm{m}$ and the Stefan-Boltzmann law must be fulfilled. If the model does not meet all mentioned criteria, the structure is iteratively adjusted, using the Henyey scheme (cf. \ref{sec:standard}). Note that this implies that a new interpolated envelope is computed, for each iteration.


\subsection{Including and calibrating the turbulent pressure} \label{sec:turb}

In convective flows, the turbulent bulk motion of the fluid contributes an additional pressure, the so-called turbulent pressure ($P_\mathrm{turb}$). This pressure is related to the convective velocity ($v_\mathrm{c}$) of the fluid by
\begin{equation}
P_\mathrm{turb} = \beta \rho v_\mathrm{c}^2. \label{eq:pturb}   
\end{equation}
Here, $\beta$ denotes a scaling factor that we will address below. In MLT, $v_\mathrm{c}$ denotes the average velocity of a fluid element \citep[eq.~7.6]{Kippenhahn}:
\begin{equation}
v_\mathrm{c}^2 = g \delta (\nabla-\nabla_\mathrm{e})\frac{\alpha_\textsc{mlt}^2H_\mathrm{P}}{8}. \label{eq:vc}
\end{equation}
Here, $\delta = -(\partial \ln \rho / \partial \ln T)_P$ {\color{black}at constant $P$}, and $H_\mathrm{P}$ denotes the pressure scale height. Furthermore, $\nabla$ and $\nabla_\mathrm{e}$ refer to the temperature gradient of the surroundings and of the convective element, respectively.

The stratification of the turbulent pressure from the interpolated $\langle \mathrm{3D} \rangle$-envelope is used above the matching point. It is again obtained using the interpolation described in Section~\ref{sec:interp}: the scaled turbulent pressure as function of the scaled thermal pressure is evaluated by interpolation in $T_\mathrm{eff}$ and $\log g$. The associated scaling factors are likewise found by interpolation and the scaling is inverted, {\color{black}i.e. the absolute quantities are computed}.

Below the matching point, we compute the turbulent pressure using Eq.~(\ref{eq:pturb}), i.e. from MLT. In order to ensure a continuous transition at the matching point, we calibrate $\beta$ in Eq.~(\ref{eq:pturb}) such that MLT recovers the turbulent pressure ($P_\mathrm{turb,m}^\mathrm{3D}$) at the base of the appended $\langle \mathrm{3D} \rangle$-envelope, i.e.
\begin{equation}
\beta = \frac{P_\mathrm{turb,m}^\mathrm{3D}}{\rho^\mathrm{1D}_\mathrm{m}(v^\mathrm{1D}_\mathrm{c,m})^2}. \label{eq:beta}
\end{equation}
{\color{black}By following this approach, we allow $\beta$ to take a different value for each iteration and hence to evolve with time, in order to ensure a continuous stratification of $P_\mathrm{turb}$. However, as discussed in Section~\ref{sec:evolution}, it turns out that $\beta$ varies little over the computed evolution from the pre-main sequences (pre-MS) to the RGB.
}

\begin{figure*}
\centering
\includegraphics[width=0.85\linewidth]{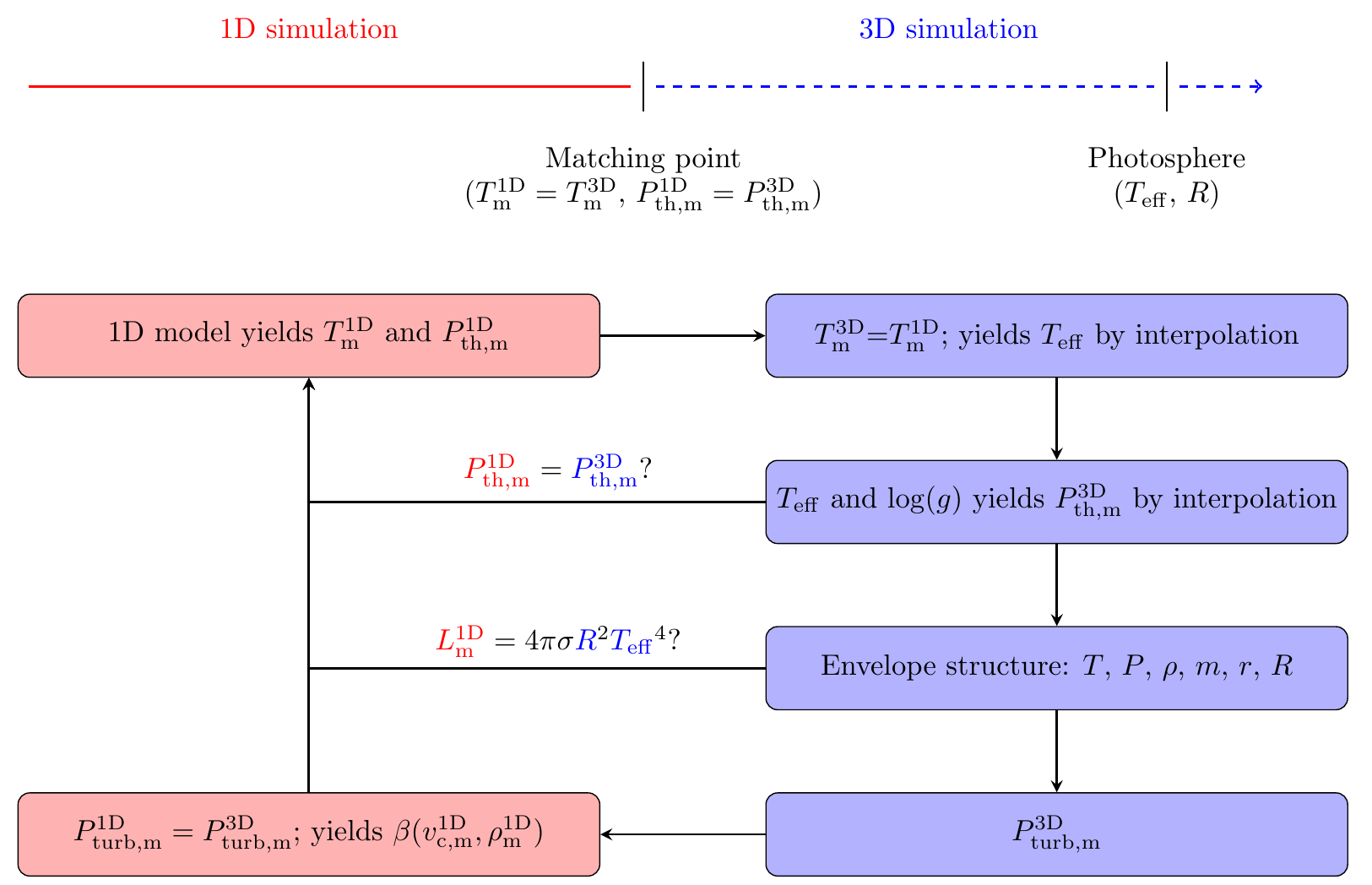}
\caption{Flowchart summarizing the coupling of 1D and 3D simulations. An interpolated $\langle \mathrm{3D} \rangle$-envelope is appended at each iteration and used as the outer boundary conditions to solve the stellar structure equations in the stellar interior. An equilibrium structure is found when the Stefan-Boltzmann law is fulfilled and a {\color{black}continuous} transition in the thermal pressure is ensured. A {\color{black}continuous} transition in the turbulent pressure is enforced by a suitable choice of $\beta$ in Eq.~(\ref{eq:pturb}).
}
\label{fig:flowchart}
\end{figure*}

The flowchart in Fig.~\ref{fig:flowchart} summarizes our implementation of $\langle \mathrm{3D} \rangle$-envelopes into \textsc{garstec}, including the calibration of the turbulent pressure.

The turbulent pressure contributes to the total pressure and must hence be accounted for, when computing the radius from the requirement of hydrostatic equilibrium also in the interior model; cf. Eq.~(\ref{eq:hydrostat}). Additionally, the turbulent pressure alters derivatives with respect to the total pressure, which enter the computation of the convective velocity, and a consistent implementation of turbulent pressure must hence iteratively take this feedback into account. Furthermore, the kinetic energy of turbulent motion also adds to the total energy, contributing with $E_\mathrm{k} = v_\mathrm{c}^2/2$ per mass unit. Its change with time constitutes a further term in the energy equation {\color{black}solved in the interior model.}. We repeat that we ignore any energy contribution from the appended envelope.


\subsection{Computing stellar oscillations} \label{sec:freqcomp}

We compute {\color{black}calibrated solar models \citep[e.g.][]{Vinyoles2017}}, requiring the models to match observations of the present-day Sun. The details of this procedure are described in Section~\ref{sec:completemodel}.
Once we have calibrated a standard solar model, we proceed to compare its structure with seismic observations. To do so, we computed model oscillation frequencies, using the Aarhus adiabatic pulsation package, \textsc{adipls} \citep{Christensen-Dalsgaard2008a}, and compare with observations from the Birmingham Solar Oscillation Network \citep[BiSON: ][]{Broomhall2009, Davies2014} in Section~\ref{sec:oscillations}.

Due to their small amplitudes, solar oscillations are well-described using perturbation theory. The perturbations are due to pressure and density fluctuations and propagate through the equilibrium structure. This is dictated by the requirement of hydrostatic equilibrium and therefore the {\em total} pressure enters the computation of stellar oscillations through Eq.~(\ref{eq:hydrostat}). The turbulent pressure itself does not enter thermodynamic quantities, such as e.g. $\Gamma_1 = \partial \ln P_\mathrm{th}/\partial \ln \rho$ or the sound speed, but its inclusion requires a revision of the linearized oscillation equations that describe stellar pulsations.

\textsc{adipls} treats stellar oscillations adiabatically. Consequently, the relative perturbation in the density ($\delta \rho/\rho$) and the relative perturbation in the thermal pressure ($\delta P_\mathrm{th}/P_\mathrm{th}$) \citep[e.g.][]{Aerts2010} are closely related:
\begin{equation}
\frac{\delta \rho}{\rho} = \frac{1}{\Gamma_1}\frac{\delta P_\mathrm{th}}{P_\mathrm{th}}
\end{equation}
This expression can be rewritten in terms of the total pressure:
\begin{equation}
\frac{\delta \rho}{\rho} = \frac{1}{\Gamma_1}\frac{P}{P_\mathrm{th}}\left( \frac{\delta P}{P} - \frac{\delta P_\mathrm{turb}}{P} \right).
\end{equation}
As argued by \cite{Houdek2017}, based on \cite{Houdek2015}, the pulsational perturbation of the turbulent pressure ($\delta P_\mathrm{turb}$) mainly contributes to the driving and damping of the oscillations, i.e. to the imaginary part of the mode frequencies. Thus, $\delta P_\mathrm{turb}$ can safely be neglected, when computing adiabatic oscillations, i.e.
\begin{equation}
\frac{\delta \rho}{\rho} \approx  \frac{1}{\Gamma^\mathrm{red}_1}\frac{\delta P}{P},
\end{equation}
where $\Gamma_1^\mathrm{red} = \Gamma_1 P_\mathrm{th}/P$. In short, turbulent pressure alters the lineraized pulsation equations by a factor of $P/P_\mathrm{th}$, which may conveniently be taken into account by including this factor in $\Gamma_1$. This is known as the reduced $\Gamma_1$ approximation \citep[cf. ][]{Rosenthal1999}.

In the case of \textsc{adipls}, the above implies that one must supply $P$, $\rho$, $r$, $\Gamma_1^\mathrm{red}$, and $q=m/M$ at each mesh point along with the  Brunt-V{\"a}is{\"a}l{\"a} frequency
\begin{equation}
\langle N^2 \rangle = \frac{1}{\Gamma_1^\mathrm{red}} \frac{\mathrm{d}\ln P}{\mathrm{d}\ln r} - \frac{\mathrm{d}\ln \rho}{\mathrm{d}\ln r}, 
\end{equation}
in order to compute stellar oscillation frequencies in a consistent manner.


\section{A complete solar model} \label{sec:completemodel}

\textsc{garstec} employs a first-order Newton iteration scheme, in order to evaluate the initial composition and the mixing length parameter that recover the global parameters of the Sun \citep{Weiss2008} at the accurately determined solar age.
Specifically, \textsc{garstec} iteratively adjusts the mixing length parameter ($\alpha_\mathrm{mlt}$) as well as the mass fraction of hydrogen ($X_\mathrm{i}$) and helium ($Y_\mathrm{i}$) on pre-MS, until the present-day relative mass fraction of heavy elements at the solar surface ($Z_\mathrm{S}/X_\mathrm{S}$), the surface radius, and the luminosity agree with observations. {\color{black}The age is set to be $4.57\,$Gyr, and the initial mass is fixed to a value that ensures that the Sun reaches a mass of $1\,$M$_\odot$ at its present age\footnote{{\color{black}\textsc{garstec} takes mass loss into account. For the present-day Sun, this mass loss depends only very weakly on exact evolution of the Sun through the HR diagram.}}}. Here, $R_\odot = 6.95508\times10^{10}\,$cm. We set $\mathrm{Z}_\mathrm{S,\odot}/\mathrm{X}_\mathrm{S,\odot}=0.01792$ based on the photospheric and meteoritic abundances {\color{black}determined by \cite{Asplund2009} {\color{black}--- that is, we use meteoritic values for all but the relevant isotopes of H, He, C, N, O, Ne, Si, S, Ar and Cr \citep[cf. ][]{Serenelli2011,Joergensen2019}}}. The Stagger grid contains a simulation of the solar surface layers for the same composition. To facilitate an easy comparison with this simulation, we set $T_\mathrm{eff,\odot}=5768.5\,$K, {\color{black}which corresponds to the effective temperature of this simulation. Moreover, for all models of the present-day Sun presented in this paper, $\log g$ deviates by less than 0.001 dex from the solar Stagger-grid simulation at the matching point.
}

Our analysis includes the following solar calibration models:

\begin{itemize}
    \item Model A ($3\mathrm{D} +P_\mathrm{turb}$) is constructed, using the method described in Section~\ref{sec:Coupling1D3D}. It hence appends an interpolated $\langle \mathrm{3D} \rangle$-envelope and calibrates $\beta$ in Eq.~(\ref{eq:pturb}) at each time-step.
    \item Model B: as model A, but the turbulent pressure has been ignored ($\beta = 0$).
    \item Model C ($\mathrm{Edd.} +P_\mathrm{turb}$) uses an Eddington grey atmosphere as the outer boundary condition (cf. Section~\ref{sec:standard}), and includes turbulent pressure. Interpolated $\langle \mathrm{3D} \rangle$-envelopes are used to calibrate $\beta$ in Eq.~(\ref{eq:pturb}) at every time-step: $\beta$ is chosen in such a way as to recover the turbulent pressure at the base of an $\langle \mathrm{3D} \rangle$-envelope with the same $T_\mathrm{eff}$ and $\log g$ as the determined 1D model. The base of the envelope is placed at a thermal pressure that is $10^{1.2}$ times higher than the thermal pressure at the density inversion of the $\langle \mathrm{3D} \rangle$-envelope.
    \item Model D is a standard solar calibration model. It employs an Eddington grey atmosphere and does not include turbulent pressure.
\end{itemize}

In accordance with \cite{Joergensen2018}, we find that the outer boundary conditions of model A-D do not alter the inferred abundance on the pre-MS: all four solar calibrations lead to similar initial compositions. The values obtained for $Y_\mathrm{i}$ and $Z_\mathrm{i}$ match within $10^{-4}$ and $10^{-5}$, respectively.

{\color{black}As discussed in Section~\ref{sec:evolution}, the evolution tracks of model A-D differ. Although model A-D only differ in the treatment of the outermost layers, one may expect this to affect the entire predicted structure of the present day Sun, despite of similar initial conditions. However, the differences in the stellar evolution tracks are rather small up until the current solar age.} Overall, we {\color{black}thus} find that the interior structure of models A-D are rather similar, despite the differences between the boundary conditions. {\color{black}
Considering the similarities in the chemical evolution, this behavior can be explained by the fact that all models are required to recover the same global parameters, i.e. $L_\odot$, $R_\odot$, $M_\odot$ and $(\mathrm{Z/X}_\odot)$, which leads to the same asymptotic entropy of the deep adiabat.
}
We will discuss the {\color{black}similarity of the structures of model A-D} further in Section~\ref{sec:soundspeed}. {\color{black}At this point, one may rather turn the argument around:}
the fact that the interior structure, including the depth of the convective envelope, is mostly unaltered by the boundary layers explains, why the initial composition is mostly insensitive to the outer boundary conditions.

The mixing length parameter ($\alpha_\textsc{mlt}$), on the other hand, differs significantly between Model A-D: the solar calibration yields a mixing length parameter of 1.78, when using standard input physics (Model D). However, when coupling 1D and 3D simulations, the calibration results in a much higher mixing length parameter --- that is, 4.88 and 3.61 for model A and B, respectively. 
As mentioned in \cite{Joergensen2018}, we find this increase in $\alpha_\textsc{mlt}$ {\color{black}to correlate with the} location of the outer boundary of the interior model. We attribute this to the fact that $\alpha_\textsc{mlt}$ must provide the entropy difference between the {\color{black}deep} asymptotic adiabat and the  outer{\color{black}most mesh point} of the interior model, determining the mean temperature gradient of the bridged region. In standard stellar models, the outer{\color{black}most mesh point of the interior model} is placed at the photosphere, and $\alpha_\textsc{mlt}$ must encompass an adequate depiction of the superadiabatic surface layers that stretch over a few thousand kilometers from the photosphere to the nearly adiabatic stellar interior. {\color{black}When appending $\langle \mathrm{3D} \rangle$-envelopes, on the other hand, the {\color{black}outermost mesh point of the interior model, i.e the} matching point, is placed within the nearly-adiabatic region below the photosphere. Consequently, $\alpha_\textsc{mlt}$ merely bridges the entropy difference within a thin layer, whilst the remaining entropy jump is prescribed by the $\langle \mathrm{3D} \rangle$-envelope.

The substantially higher value of the mixing length parameter for model A and B illustrates the advantages of our coupling procedure over post-evolutionary patching (cf.~Section~\ref{sec:intro}), where 3D simulations are only used to substitute the outermost layers of the final model. In our coupling procedure, the interior model is required to match the physical conditions at the base of the appended envelope. The mixing length parameter ensures this by recovering the correct entropy difference between the deep adiabat and the matching point. For patched models, on the other hand, no such conditions are imposed, and patching an $\langle \mathrm{3D} \rangle$-envelope to a SSM, such as model D, therefore leads to a structure that shows discontinuities in the stratification of several physical quantities. For a detailed discussion of this issue, we refer to \cite{Joergensen2017}.
}

The dependence of  $\alpha_\textsc{mlt}$ on the location of the boundary conditions hence advocates the use of a depth dependent mixing length parameter in MLT. This is also found by \cite{Joergensen2018}, according to whom the value of $\alpha_\textsc{mlt}$ that is obtained for the coupled models depends on the depth of the matching point.
A single value for $\alpha_\textsc{mlt}$ is insufficient to encompass the complexity of superadiabatic convection.
We note that the benefits and implications of varying the mixing length parameter with depth have been discussed by \cite{Schlattl1997} and \cite{Trampedach2011}.

Taking turbulent pressure into account likewise affects $\alpha_\textsc{mlt}$.  The inclusion of turbulent pressure affects $\alpha_\textsc{mlt}$ through a complex feedback\footnote{As an example of the feedback on the stellar structure, we note that it is important to iteratively account for turbulent pressure, when computing the convective velocity that enters the turbulent pressure. This was already mentioned in Section~\ref{sec:turb}. If this feedback is ignored, $\alpha_\textsc{mlt} = 1.6514$ for Model C. The initial composition is unaffected.}, as it directly or indirectly affects different quantities. For instance, since turbulent pressure enters the hydrostatic pressure, {\color{black}it also} affects the stellar radius, which must be accounted for by a change in $\alpha_\textsc{mlt}$.

As regards Eq.~(\ref{eq:pturb}), we find that $\beta_\odot \approx 1$ for Model A, i.e. when appending interpolated $\langle \mathrm{3D} \rangle$-envelopes at every time-step. Here, $\beta_\odot$ is the value of $\beta$ for our solar calibration models, i.e. at the age of the present Sun. The time evolution of $\beta$ will be addressed in Section~\ref{sec:evolution}. Using an Eddingon grey atmosphere and MLT, on the other hand, $\beta_\odot \approx 2$ (Model C). This reflects the associated change in the calibrated mixing length parameter through the dependence of $v_\mathrm{c}$ on $\alpha_\textsc{mlt}$ (cf. Eq.~\ref{eq:vc}). 

The initial composition, $\alpha_\textsc{mlt}$, and $\beta_\odot$ are summarized in Table~\ref{tab:param} for Model A-D. 

\begin{table}
	\centering
	\caption{Calibration values for the initial solar composition, the mixing length parameter and the scaling factor of the turbulent pressure. The latter is given at the age of the present Sun.}
	\label{tab:param}
	\begin{tabular}{lccccccccccccccccccccc} 
		\hline
		Model & $Y_\mathrm{i}$ & $Z_\mathrm{i}$ & $\alpha_\textsc{mlt}$ & $\beta_\odot$ \\
	    \hline
        A ($3\mathrm{D}+P_\mathrm{turb}$) & 0.2636 & 0.0149 & 4.8760 & 0.9977 \\
        B ($3\mathrm{D}$) & 0.2637 & 0.0149 & 3.6147 & --- \\
        C ($\mathrm{Edd.}+P_\mathrm{turb}$) & 0.2638 & 0.0149 & 1.7362 & 1.9478 \\
	    D ($\mathrm{Edd.}$) & 0.2638 & 0.0149 & 1.7824 & --- \\
	    \hline
\end{tabular}
\end{table}


\subsection{The turbulent pressure profile} \label{sec:turbprofile}

\begin{figure}
\centering
\includegraphics[width=0.85\linewidth]{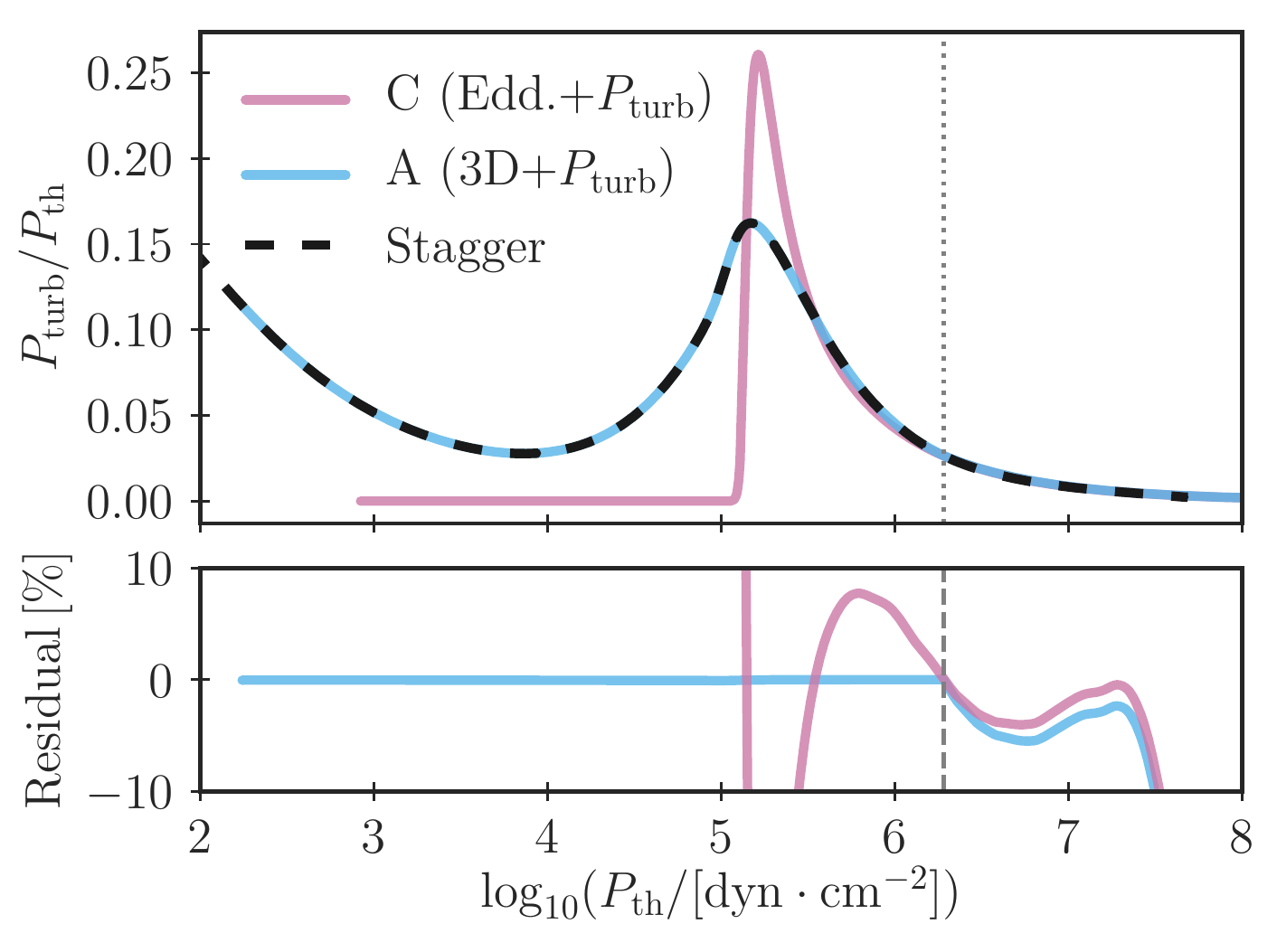}
\caption{Ratio between turbulent pressure and thermal pressure in \textsc{garstec} models of the present Sun. We include model A ($3\mathrm{D} +P_\mathrm{turb}$) as well as model C ($\mathrm{Edd.} +P_\mathrm{turb}$) described in Section~\ref{sec:completemodel}. The dotted grey line indicates the position of the mesh point, at which the scaling factor of the turbulent pressure, i.e. $\beta$, is calibrated. {\color{black}In model A, this corresponds to the mesh point beyond which we append an interpolated $\langle \mathrm{3D}\rangle$-envelope --- i.e. it is the matching point}. For comparison, we have included the original Stagger-grid simulation of a solar envelope.
}
\label{fig:Ptsun}
\end{figure}

Our {\color{black}interpolation scheme} is able to reliably recover the stratification of the turbulent pressure in the Stagger-grid simulations. Furthermore, our {\color{black}coupling scheme} ensures a continuous transition in turbulent pressure between the appended $\langle \mathrm{3D} \rangle$-envelope and the interior model. This is illustrated in Fig.~\ref{fig:Ptsun}. As can be seen from the figure, the residuals between turbulent pressure of model A relative to its thermal pressure and the corresponding quantity for the original $\langle \mathrm{3D} \rangle$-envelope are negligible at any depth above the matching point. Moreover, by construction, both model A and C recover the correct turbulent pressure at the matching point.

As regards Model C, we note that MLT recovers the correct stratification of the turbulent pressure at high thermal pressure, when setting $\beta_\odot \approx 2$. At low thermal pressure, $v_\mathrm{c}$ is zero {\color{black}because the appended Eddington atmosphere is, by definition, radiative.}

\begin{figure}
\centering
\includegraphics[width=0.85\linewidth]{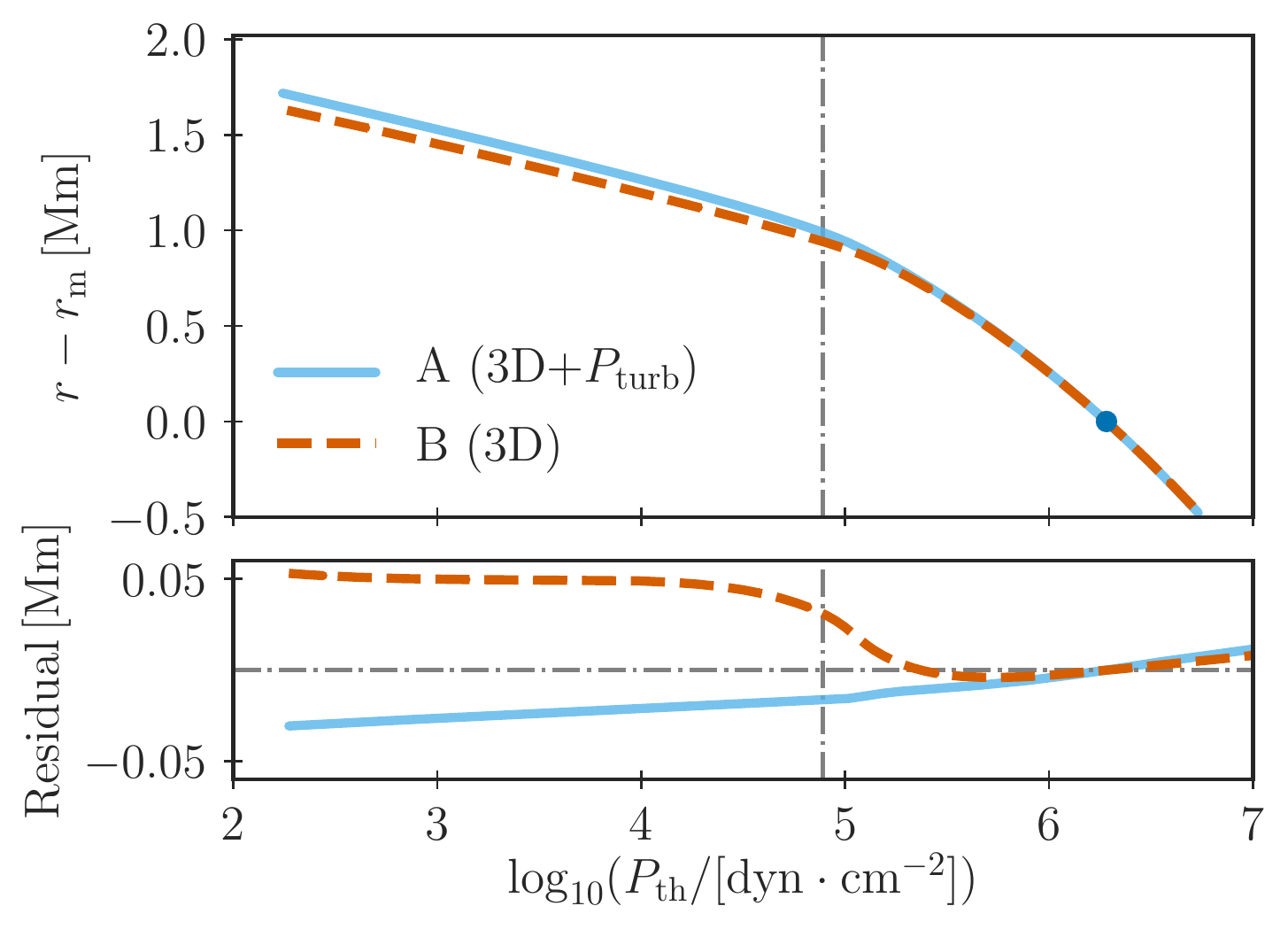}
\caption{\textbf{Upper panel:} height above the base ($r_\mathrm{m}$) of the appended $\langle \mathrm{3D} \rangle$-envelope for model A ($3\mathrm{D} +P_\mathrm{turb}$) and B ($3\mathrm{D}$). The location of the base of the envelope, i.e. the matching point, is indicated with a blue dot. The {\color{black}dash-dotted} grey line denotes the location of the stellar surface, where the Stefan-Boltzmann law is required to be fulfilled. \textbf{Lower panel:} The associated residuals between the original solar Stagger-grid envelope and our solar calibration models. Positive residuals imply that the outermost layers of the \textsc{garstec} model are less extended than the original Stagger-grid envelope.
{\color{black}As discussed in Section~\ref{sec:turbprofile}, we mainly attribute the discrepancy between model A and the Stagger-grid simulation to deviations in the density stratification.}
}
\label{fig:rvsp}
\end{figure}

Adding turbulent pressure leads to an additional pressure-gradient force that counteracts gravity and hence affects the radius attributed to each mesh point of the model. Fig.~\ref{fig:rvsp} shows that the implementation of turbulent pressure leads to a better agreement between the depth scale of the underlying Stagger-grid simulation and the inferred radius of each mesh point in our \textsc{garstec} models. For Model A, the accumulated residual over the entire envelope is reduced 
to 30 km, {\color{black}i.e. $0.03\,$Mm}. The inferred envelope of Model A is thus larger than the original Stagger-grid envelope. Without turbulent pressure the residual is 50 km, {\color{black}i.e. $0.05\,$Mm}, and has the opposite sign (Model B). 

\begin{figure}
\centering
\includegraphics[width=0.85\linewidth]{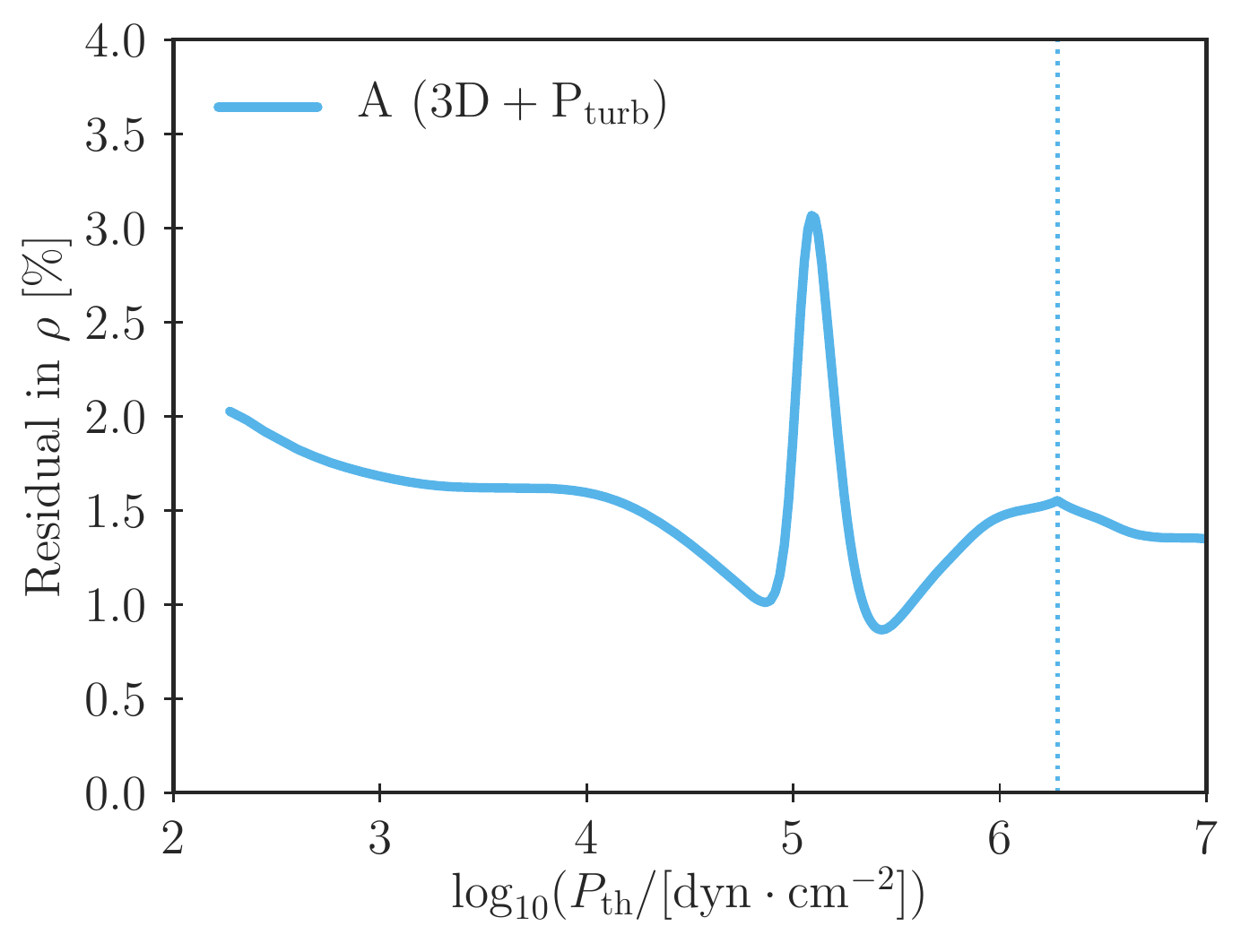}
\caption{\color{black}Residuals between the density stratification of the original solar Stagger-grid envelope and model A. Positive residuals indicate that the density of model A is too low. Different contributions to these residuals are discussed in the text.
}
\label{fig:resrho}
\end{figure}

{\color{black}The radius that is attributed to each mesh point follows from the requirement of hydrostatic equilibrium, i.e. Eq.~(\ref{eq:hydrostat}). The residuals in $r$ hence reflect discrepancies in either $P$ or $\rho$.}
{\color{black}For model A,} the pressure is correctly recovered, and the remaining residual in $r$ is due to the fact that the density inferred by the 1D stellar evolution code is systematically 1-3 per cent too low throughout the appended $\langle \mathrm{3D} \rangle$-envelope \citep[cf. ][]{Joergensen2018}. {\color{black}This is illustrated in Fig.~\ref{fig:resrho}}. The inferred radius depends on the density through Eq.~(\ref{eq:hydrostat}): $\mathrm{d}r\propto \rho^{-1}$. A too low density hence leads to a too large radius.

The reason for this systematic deviation of the density is many-fold: {\color{black}firstly, the largest deviation occurs close to the density inversion near the surface, where the temperature changes rapidly. In this region, even small interpolation errors may lead to sizeable discrepancies in the temperature{\color{black}, when computing the temperature stratification of the appended $\langle \mathrm{3D} \rangle$-envelope using the interpolation scheme by \citet{Joergensen2017}. Such interpolation errors in the temperature stratification translate} into errors in the density through the EOS (cf. Mosumgaard et al., in prep.). However, whilst the temperature residuals {\color{black}between the interpolated 3D envelope and the solar Stagger-grid simulation} reach $0.05$ per cent at the density inversion, {\color{black}these temperature} residuals are negligible throughout the rest of the $\langle \mathrm{3D} \rangle$-envelope. Thus, only the spike close to $\log_\mathrm{10}(P_\mathrm{th}) \approx 5$ may partly be attributed to interpolation errors. Secondly, the 1D and 3D simulation may deviate slightly in composition and hence in the mean molecular weight. However, we have attempted to reduce this contribution as far as possible by ensuring a large degree of consistency between the Stagger grid and the input physics in \textsc{garstec} (cf. Section~\ref{sec:standard}). It is also worth stressing that the envelope is fully mixed. Thus, there are no variations in the compositions that may lead to spikes or similar features.

Thirdly,} the 1D and 3D simulations do not employ the same EOS. {\color{black}Finally}, the density is a non-linear function of the temperature, the pressure and the chemical composition. Consequently, {\color{black}for the composition of a given model}, the average density of a 3D simulation is not the same as the density derived from the mean of the underlying quantities, i.e. {\color{black}$\rho (\langle P \rangle,\langle T \rangle) \neq \langle \rho (P,T) \rangle$} (cf.~Collet, private communications).

{\color{black}In the case of model B, the density as function thermal pressure is very similar to that of model A. On its own, this would result in the radius of model B being too large. However, the total pressure of model B is lower than the pressure of the solar Stagger-grid simulation, due to the neglect of turbulent pressure. All in all, this results in $r$ being too low: according to Eq.~(\ref{eq:hydrostat}), $\mathrm{d}r\propto P$. The extent to which the neglect of the turbulent pressure affects the residuals in $r$ can be seen by comparing Fig.~\ref{fig:Ptsun} with Fig.~\ref{fig:rvsp}: the strong increase in the residuals around $\log_\mathrm{10}(P_\mathrm{th}) \approx 5$ coincides with the peak in $P_\mathrm{turb}/P_\mathrm{th}$, i.e. the region where the contribution of the turbulent pressure increases by an order of magnitude.
}

\subsection{The sound speed profile} \label{sec:soundspeed}

\begin{figure*}
\centering
\includegraphics[width=0.85\linewidth]{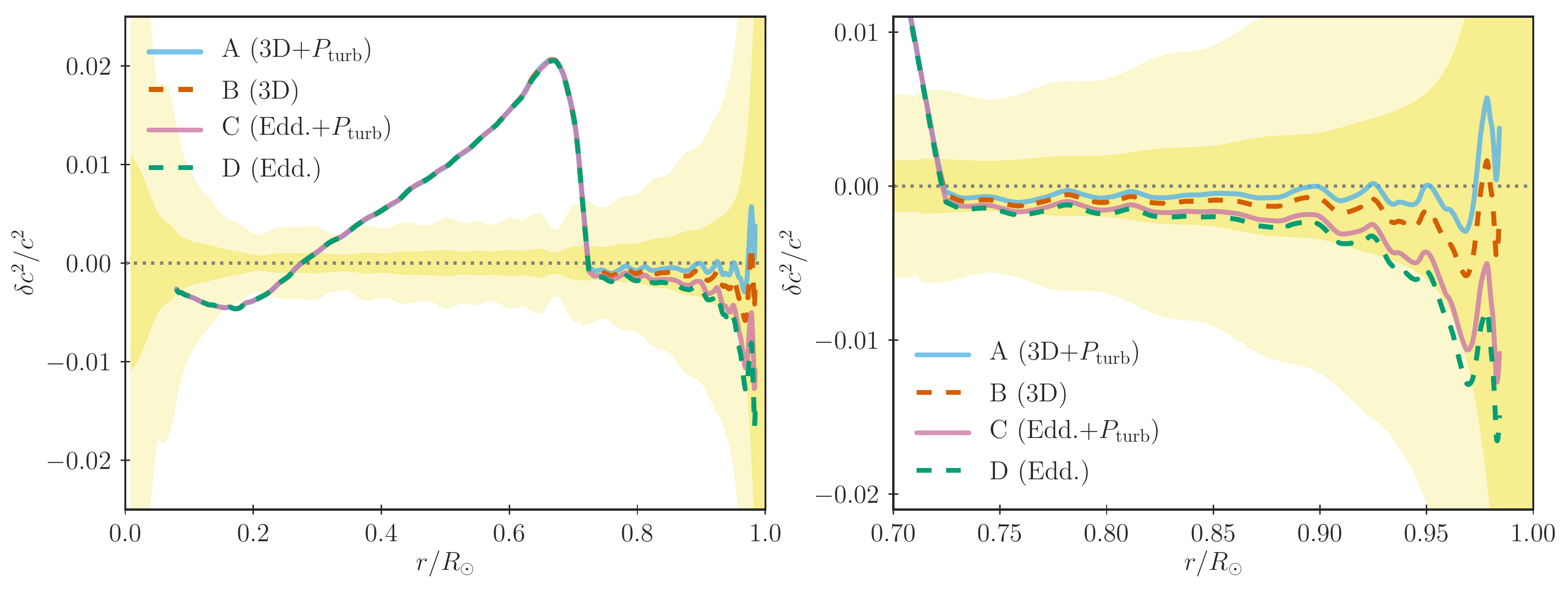}
\caption{\textbf{Left panel:} Squared sound speed difference between models A-D and the Sun. The shaded yellow area shows the statistical and conservative error estimates by Degl'Innocenti et al. (1997). {\color{black}These errors are dominated by the error that stems from the choice of reference model but also include observational errors --- see text, for further details. We use the reference model by Basu \& Antia (2008).} 
The base of the $\langle \mathrm{3D} \rangle$-envelope is located approximately 1 Mm below the photosphere, which roughly corresponds to $1.5\times 10^{-3} R_\odot$. \textbf{Right panel:} zoom-in in convective {\color{black}zone. Although the depth of the appended $\langle \mathrm{3D} \rangle$-envelope is only $1.5\times 10^{-3} R_\odot$, and although all models reach the same deep asymptotic adiabat, the different outer boundary conditions of model A-D clearly affect the stratification of the convection zone far beyond the extent of the $\langle \mathrm{3D} \rangle$-envelope.}
}
\label{fig:ccomp}
\end{figure*}

In order to assess the extent to which the evaluated solar structure is affected by the outer boundary conditions, we have computed the sound speed ($c$) of each of our four solar calibration models (Model A-D):
\begin{equation}
c^2 = \frac{\Gamma_1 P_\mathrm{th}}{\rho}.    
\end{equation}
The sound speed is an essential quantity, because it determines the propagation of acoustic oscillations in the solar interior. Conversely, the actual sound speed in the Sun can be reconstructed from solar oscillations by the means of inversion. This is commonly {\color{black}done} with respect to a reference model \citep[e.g. ][]{Pijpers1992, jcd1995}. In this paper, we employ a solar sound speed profile that was inferred by \cite{Basu2008}, using such inversion techniques. For each solar calibration, we computed the difference between the squared sound speed of the model and the Sun:
\begin{equation}
\frac{\delta c^2}{c^2} = \frac{c^2_\mathrm{sun}-c^2_\mathrm{mod}}{c^2_\mathrm{sun}}
\end{equation}
Here, $c_\mathrm{mod}$ and $c_\mathrm{sun}$ denote the sound speed of the model and the Sun, respectively. The results are shown in Fig.~\ref{fig:ccomp}. {\color{black}}For comparison, we include the uncertainties by \cite{Degl1997}. {\color{black}These uncertainties include observational errors on the measured p-mode frequencies, uncertainties of the inversion method and errors that stems from the choice of reference model. The resulting uncertainties on the sound speed profile are mostly dominated by the latter contribution \citep[cf. Fig.~3 in ][]{Vinyoles2017}. In short, \cite{Degl1997} use a small set of reference models with different input physics and vary the parameters that enter the inversion method. By adding the squared variations of the sound speed that result from this exploration of the associated parameter space, \cite{Degl1997} arrive at an estimate of $\delta c/c$. They refer to this estimate as the statistical error. Noting that their exploration of the parameter space may not have been exhaustive, they furthermore give an alternative estimate of $\delta c/c$ by adding up all errors linearly. They denote this error as being conservative.
}

As can be seen from Fig.~\ref{fig:ccomp}, we find that the inferred stratification of the outer convective layers is affected by the different boundary conditions of models A-D. Indeed, the coupling of 1D and 3D simulations leads to a better agreement between model predictions and the observationally inferred solar sound speed profile. In this connection, the inclusion of turbulent pressure plays a minor role.

However, the sound speed profile of the deep interior stays unaltered. Thus, the models A-D agree on the structure of the radiative zone, which explains why the initial composition is only marginally affected by the boundary conditions (cf. Section~\ref{sec:Coupling1D3D}). 
In particular, all four solar calibration models agree on the location of the base of the convective envelope. We define the base of the convective envelope ($r_\mathrm{cz}/R_\odot$) as usual by application of the Schwarzschild criterion for convective instability. For models A-D, $r_\mathrm{cz}/R_\odot$ is 0.7242, 0.7242, 0.7245, and 0.7245, respectively. According to \cite{Basu1997}, an inversion of solar oscillation frequencies suggests that the base of the convection zone is located at a radius of $0.713\pm 0.001 \,\mathrm{R}_\odot$. This is a well-known disagreement of solar models using the more recent lower present-day metallicities, in particular those by \citet{Asplund2009}. The improvements of the outer boundary conditions that results from the coupling 1D and 3D
have little effect on the location of the bottom of the convective envelope.

There is also a large anomaly in $\delta c^2/c^2$ close to the base of the convective envelope (Fig.~\ref{fig:ccomp}). This so-called tachocline anomaly is likewise a well-known shortcoming of state-of-the-art solar models \citep{jcd1985,jcd1988}, and again the use of AGSS09 abundances is found to lead to a particularly strong disagreement with observations \citep{Serenelli2009}. 


\subsection{Solar oscillations} \label{sec:oscillations}

\begin{figure}
\centering
\includegraphics[width=0.85\linewidth]{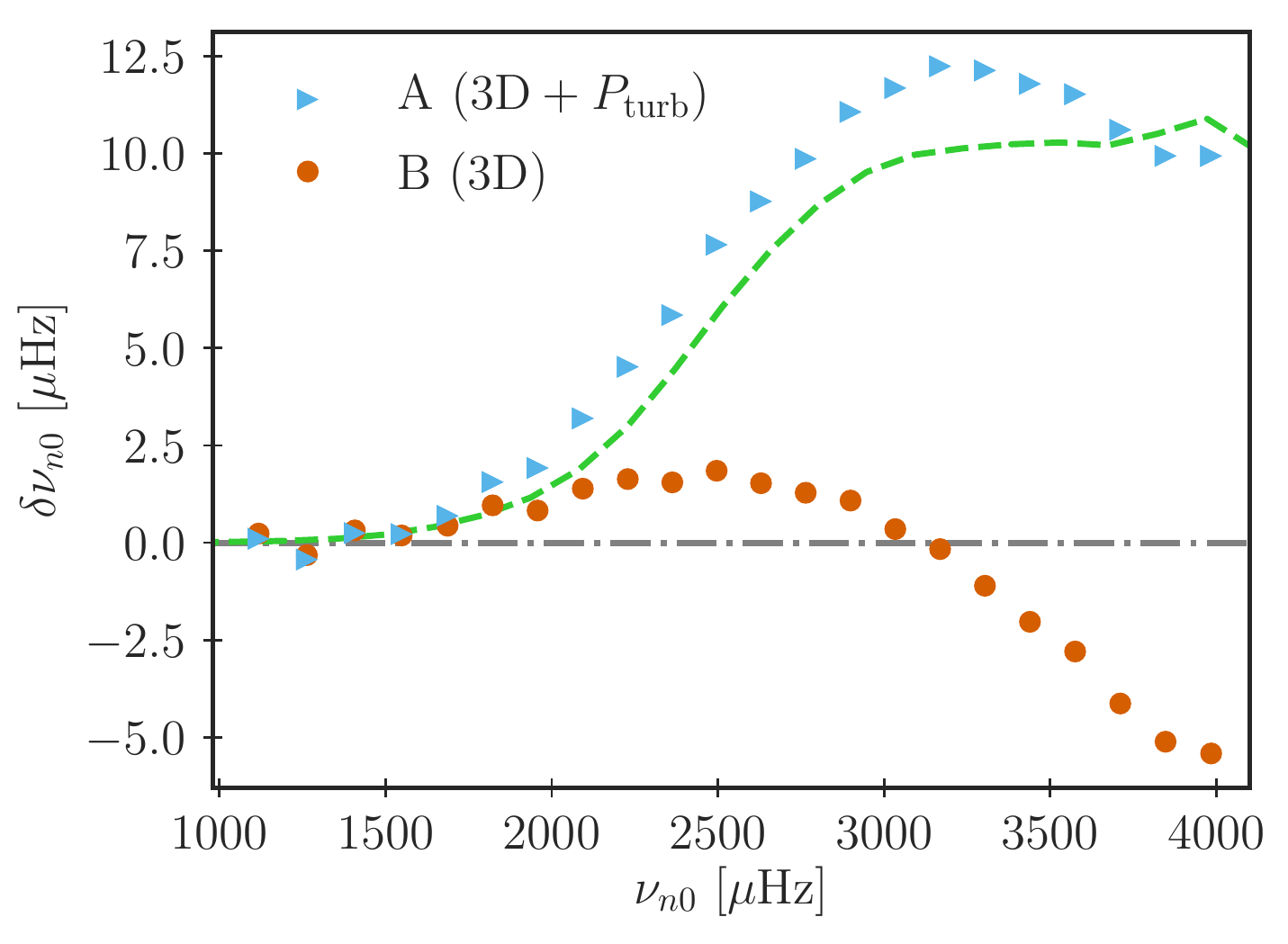}
\caption{Frequency difference between BiSON observations and a model of the present Sun that appends an interpolated $\langle \mathrm{3D} \rangle$-envelope and calibrates $\beta$ at each time-step (model A). Model B is included for comparison, to demonstrate the effect of including turbulent pressure. The dashed light green line shows the modal effect (with opposite sign) that was originally published in Fig.~1 in Houdek et al. (2017). The modal effect and the structural effect combine to an effective residual of at most $2\,\mu$Hz everywhere. The modal effect was kindly provided by G.~Houdek. Only radial modes ($\ell=0$) with frequencies above $1000\,\mu$Hz are included.
}
\label{fig:Frek_G3D}
\end{figure}

Following the approach outlined in Section~\ref{sec:freqcomp}, we computed adiabatic model frequencies for model A. In order to assess the extent, to which our method corrects for the structural contribution to the surface effect, we compared these model frequencies to BiSON observations. The comparison is shown in Fig~\ref{fig:Frek_G3D}, where 
\begin{equation}
\delta \nu_{n \ell} = \nu^\mathrm{obs}_{n \ell}-\nu^\mathrm{mod}_{n \ell}.     
\end{equation}
Here, we let $\nu_{n\ell}$ denote the mode frequency with radial order $n$ and degree $\ell$. The superscripts 'obs' and 'mod' refer to the observed and the model frequencies, respectively. $\delta \nu_{n \ell}$ denotes the remaining frequency residual. For simplicity, we only include radial modes ($\ell=0$). We hence avoid the necessity of scaling $\delta \nu_{n \ell}$ by the ratio between the mode {\color{black}inertia (or modal mass)} and the mode {\color{black}inertia} of a radial mode with the same frequency \citep[cf. ][]{1986ASIC..169...23C}. 

We find that the adiabatic model frequencies of model A are systematically lower than the observed frequencies. The frequency residual is close to $12\,\mu$Hz in the vicinity of the frequency of maximum power ($v_\mathrm{max,\odot}=3090\,\mu$Hz). Overall, the obtained frequency residuals are in very good agreement with the structural surface effect inferred by \cite{Houdek2017} and \cite{Joergensen2018} using post-evolutionary patching.

\cite{Houdek2017} have determined the modal contribution to the surface effect of the Sun, using a time-dependent, non-local treatment of pulsational perturbations. Such effects include non-adiabatic energetics. According to \cite{Houdek2017}, the modal surface effect counteracts the structural surface effect. Furthermore, the modal surface effect matches the frequency residuals of model A as a function of frequency \citep[cf. Fig.~1 and 5 in ][]{Houdek2017}. 
To illustrate this, we have included the modal surface effect from \cite{Houdek2017} in Fig.~\ref{fig:Frek_G3D} --- with the opposite sign, in order to facilitate an easy comparison with the structural surface effect. 
The remaining residuals between observations and the model frequencies from model A are at most $2\,\mu$Hz for frequencies below $4000\,\mu$Hz,
once modal effects have been subtracted.
Physical differences between our solar calibration models and those used by \cite{Houdek2017} may partly contribute to the remaining frequency residuals. However, we note that \cite{Houdek2017} obtain the same deviation of at most $2\,\mu$Hz.

From the above, we conclude that model A successfully eliminates the structural surface effect, leaving only modal effects to be corrected for. Our method hence overcomes the structural shortcomings of the surface layers that haunt current standard stellar structure models. The remaining systematic offset between model frequencies and observations is mainly due to the assumption of adiabaticity that enters the frequency computation.


\subsection{Evolution of the Sun} \label{sec:evolution}

\begin{figure}
\centering
\includegraphics[width=0.85\linewidth]{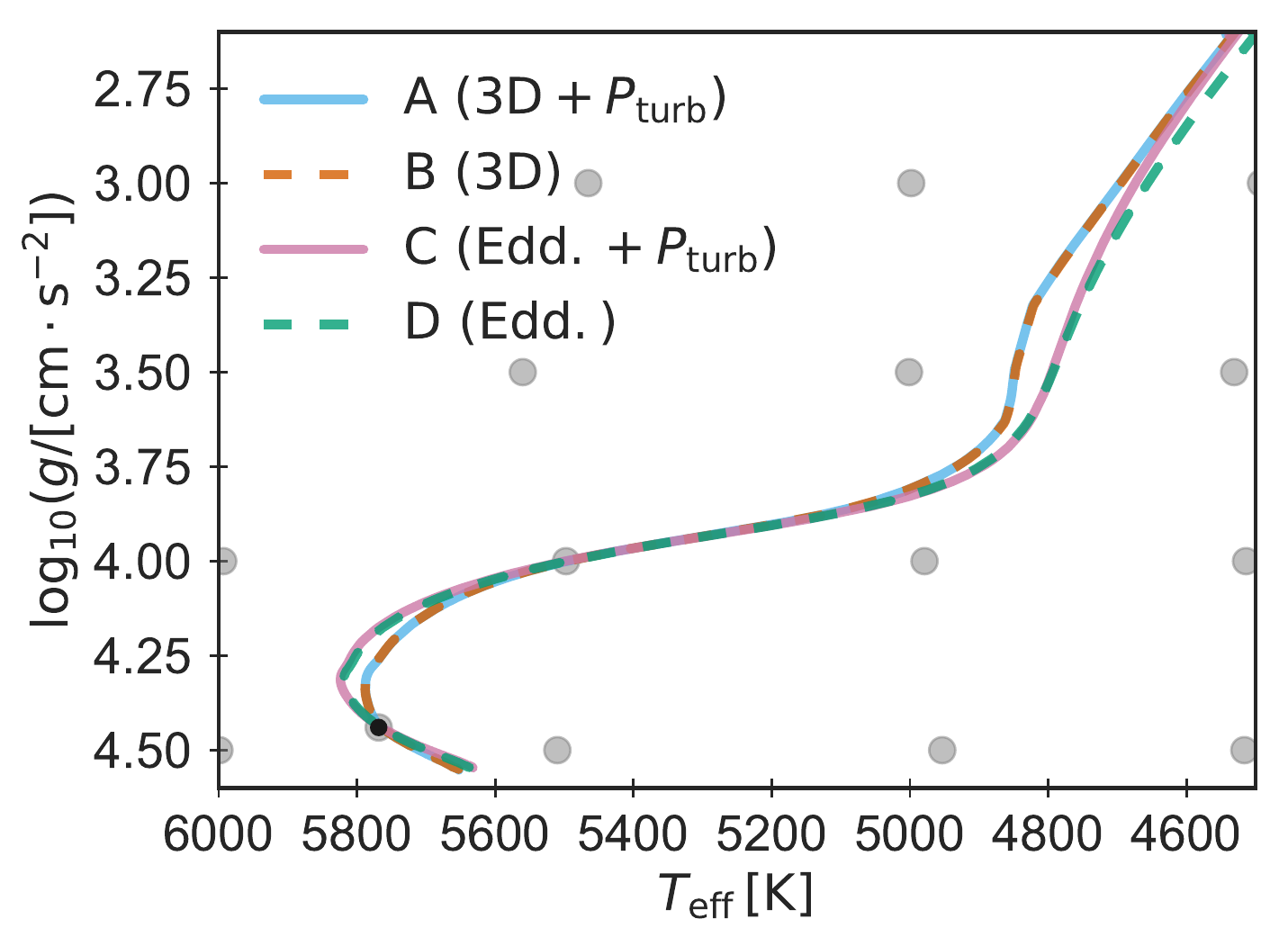}
\caption{Kiel diagram showing the evolutionary tracks of four different solar models corresponding to model A-D, starting from the zero age main sequence (ZAMS). The grey circles show the location of a selection of Stagger-grid simulations that entered the interpolation. The black dot indicates the position of the present Sun.
}
\label{fig:paths_sun}
\end{figure}

Below, we discuss the predicted evolution of the Sun based on our four calibration models of the present Sun (model A-D). The computed evolutionary tracks are shown in Fig.~\ref{fig:paths_sun}.

We find that the coupling of 1D and 3D simulations leads to a lower effective temperature at the turn-off point, i.e. at the end of the MS, and to higher effective temperatures on the RGB. This is consistent with the findings of \cite{Mosumgaard2018}, who use a variable $\alpha_\textsc{mlt}$ and altered boundary conditions in order to encapsulate the basic properties of 3D simulations \citep{Trampedach2014a,Trampedach2014b}: \cite{Mosumgaard2018} conclude that their implementation of information from 3D simulations lead to higher temperature on the RGB. A detailed comparison with this method {\color{black}will be discussed in Mosumgaard et al. (in preparation).} 

As can be seen from Fig.~\ref{fig:paths_sun}, the evolution of solar models with Eddington atmospheres (model C and D) are not significantly altered by taking turbulent pressure into account until the RGB. When coupling 1D and 3D simulations, turbulent pressure is mostly irrelevant for the evaluated evolutionary track.

We note that models that include $\langle \mathrm{3D} \rangle$-envelopes on the fly are subject to interpolation errors, due to the low number and the distribution of the simulations in the Stagger grid. This can be seen from Fig.~\ref{fig:paths_sun}, where kinks in the evolutionary tracks reflects the triangulation that underlies the interpolation procedure and thereby indirectly reflects the depth of the appended $\langle \mathrm{3D} \rangle$-envelope. This issue is discussed in detail in the companion paper of \cite{Joergensen2018}, and we will hence not touch further upon it in this paper. We solely note that our results call for a refinement of the Stagger grid.

\begin{figure}
\centering
\includegraphics[width=0.85\linewidth]{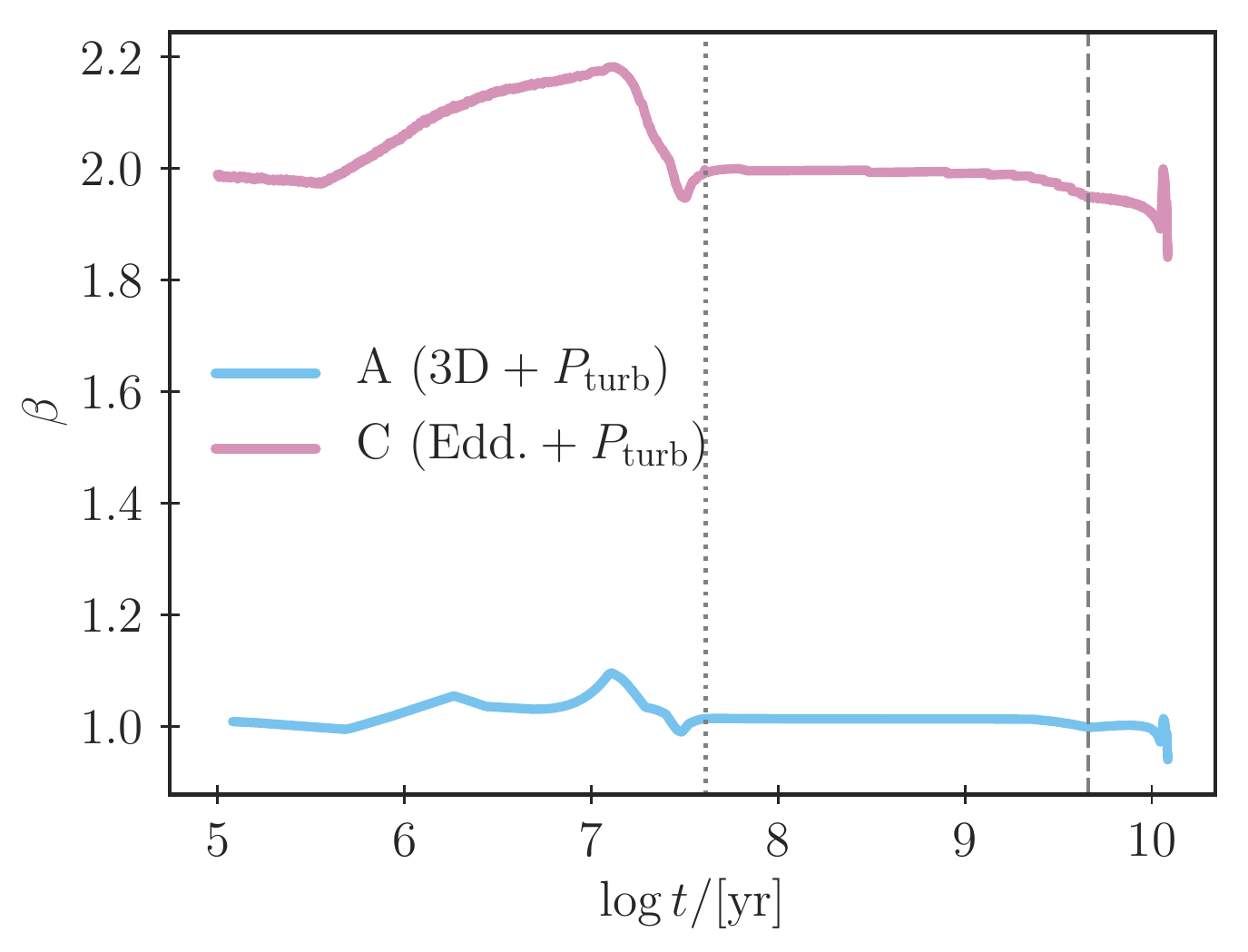}
\caption{The calibrated scaling factor ($\beta$) of the turbulent pressure as function of the stellar age for model A ($3\mathrm{D} +P_\mathrm{turb}$) and model C ($\mathrm{Edd.} +P_\mathrm{turb}$). The dotted line indicates the age of the models on the ZAMS, whilst the dashed line marks the age of the current Sun. We have computed the stellar evolution up until an age of roughly $12\,$Gyr.
}
\label{fig:beta_sun}
\end{figure}

From our analysis we conclude that $\beta$ in Eq.~(\ref{eq:pturb}) is approximately constant from the pre-MS to the RGB {\color{black}--- that is, at least in the solar case}. This result is summarized in Fig.~\ref{fig:beta_sun}.

The fact that $\beta$ is constant as a function of time partly reflects the fact that we have assumed $\alpha_\textsc{mlt}$ to be constant across the HR diagram, i.e. to correspond to the value obtained from the respective solar calibration.
This is a simplifying approximation: asteroseismic and spectroscopic measurements and simulations point towards a variation of $\alpha_\textsc{mlt}$ with $T_\mathrm{eff}$, $\log g$ and $\mathrm{[Fe/H]}$ \citep[see for example][]{Trampedach2014b,Magic2015,Tayar2017}. However, no consensus has yet been reached. Also, this simplifying assumption is commonly used, and so by employing a constant mixing length parameter, we facilitate a more straight-forward comparison with standard stellar models.

Whilst the use of a constant $\alpha_\textsc{mlt}$ may affect the predicted evolution after the MS, we furthermore note that this assumption is irrelevant for the solar calibrations themselves: the variation in $\alpha_\textsc{mlt}$ across the relevant parameter space is negligible (cf. figs~4 and 5 in \citealt{Trampedach2014b}).


\section{Conclusions}

In previous papers, we have discussed how to  improve upon the modelling of stars with convective envelopes by replacing the outermost superadiabatic layers in 1D stellar evolution codes with mean 3D structures \citep{Joergensen2017,Joergensen2018}. Building upon these previous steps, we have now included turbulent pressure at every time-step of the predicted stellar evolution based on 3D RHD simulations. We have shown that our improved method reliably recovers the mean stratification of 3D hydrodynamic simulations of stellar envelopes, in the case of the present Sun. Furthermore, our method
removes most of the need for describing convection by MLT in the superadiabatic regions near the surface. The appended 3D boundary layers are coupled to the nearly adiabatic regions during the entire stellar evolution.

We have applied our method to the solar case and did a complete standard solar model calibration for different combinations of atmosphere model and turbulent pressure inclusion. 
By comparing to helioseismic observations, we show that our method fully corrects for the structural contribution to the surface effect. Only modal effects are still left to be accounted for, but the remaining discrepancy agrees with the predicted correction if oscillations frequencies would be computed including deviations from adiabaticity \citep{Houdek2017}. {\color{black}The computation of oscillation frequencies, however, is not an issue of hydrostatic stellar models, but concerns pulsation codes.}  Furthermore, we show that our method leads to a better reconstruction of the solar sound speed profile than the standard procedure does. However, we note that this conclusion is not necessarily compelling, due to the large error-bars involved.

We find that the altered boundary layers affect the predicted solar evolution: our method leads to higher effective temperatures on the RGB and shifts the turn-off point at the end of the MS. The consistent implementation of turbulent pressure, however, is found to have a negligible effect on the evolution track. 

The described implications for stellar evolution tracks will have ramifications for the asteroseimically and spectroscopically inferred global parameters of low-mass stars with convective envelopes. This is relevant for the interpretation of high-quality data from the CoRoT, \textit{Kepler}, K2, and TESS space missions as well as for the upcoming PLATO mission \citep[e.g.][]{Ricker2015,Miglio2017}: in its search for habitable Earth-like planets, PLATO will focus on solar-like dwarfs. We will address the implications for stellar parameter estimates in a forthcoming paper.

Our analysis shows how essential a realistic treatment of superadiabatic convection is for the computation and applicability of stellar models. In the light of this, our method constitutes a useful improvement of 1D stellar evolution codes. However, we note that our method is presently subject to interpolation errors due to the low number of existing 3D simulations. These errors may potentially alter the predicted evolution, affecting the inferred stellar parameter estimates. The necessity to remedy this issue calls for refinement of current grids of 3D envelopes, since the robustness of the interpolation depends on how densely the available 3D simulations cover the relevant regions of the parameter space. For a more detailed discussion on interpolation errors and the implications of the coupling of 1D and 3D simulations on the stellar evolution tracks, we refer to our forthcoming paper (Mosumgaard et al., in prep.).

\section*{Acknowledgements}
We record our gratitude to J.~Christensen-Dalsgaard, R.~Collet and Z.~Magic for their collaboration. We furthermore thank G.~Houdek for kindly providing the modal corrections that are presented in Figure~\ref{fig:Frek_G3D} of this paper and an anonymous referee for a careful reading and helpful comments.




\bibliographystyle{mnras}
\bibliography{manual_refs,mendeley_export}








\bsp	
\label{lastpage}
\end{document}